\documentclass[11pt,a4paper]{article}
\usepackage{jcappub}

\def\be{\begin{equation}}
\def\ee{\end{equation}}
\def\beq{\begin{equation}}
\def\eeq{\end{equation}}
\def\bea{\begin{eqnarray}}
\def\eea{\end{eqnarray}}
\def\bml{\begin{subequations}}
\def\blea{\bml\begin{eqnarray}}
\def\elea{\end{eqnarray}\end{subequations}}

\begin{document}

\title{Kinetic Theory and Hydrodynamics of Cosmic Strings}

\author{Vitaly Vanchurin}

\emailAdd{vvanchur@umn.edu}

\date{\today}

\affiliation{Department of Physics, University of Minnesota, Duluth, Minnesota, 55812}

\abstract{
We develop further a kinetic theory of strings and derive a transport equation for a network of cosmic strings with Nambu-Goto evolution, interactions and background gravitational effects taken into account. We prove an H-theorem and obtain necessary and sufficient conditions for a thermodynamic equilibrium. At the lowest order the equilibrium is estimated by the von Mises-Fisher distributions parametrized by mean directions and dispersions of the right- and left-moving tangent vectors. Under assumption of a local equilibrium we derive a complete set of hydrodynamic equations that govern the evolution of strings on large scales. We also argue that on small scales the assumption of a local equilibrium would break down, and non-equilibrium steady states, described by the Sinai-Ruelle-Bowen distributions, should be used instead. }

\maketitle

\section{Introduction}

Cosmic strings are predicted by many models of symmetry breaking phase transitions \cite{PhaseTransitions} and give rise to very distinct and detectable signatures such as gravitational lensing \cite{Lensing}, CMB non-Gaussianities \cite{Nongaussianity}, gravitational waves \cite{Gravitywaves}, ultrahigh energy cosmic rays \cite{UHECR}, radio signals \cite{Radio} etc. It is also believed that cosmic superstrings can be formed at the end of brane inflation \cite{BraneInflation} which opens a possibility of testing the models of string theory in the cosmological settings. However, all of the phenomenological predictions are based on the statistical properties of cosmic strings, which remain largely unexplored.

In the early work on strings statistics \cite{OldAnalytical}, it was expected that a typical length of closed loops would scale linearly with time, but some old numerical simulations showed no evidence for such scaling \cite{OldTurok, OldBB, BB, OldAS, OldNumerical}. The average distance between long strings was observed to scale linearly with time, but the size of the smallest wiggles as well as the size of loops remained at the resolution scale of the simulations. More recently, the high-precision numerical simulations \cite{NewNumerical} with much finer resolution were developed to demonstrate that the long strings do obey a scaling law on a wide range of scales described by a universal power spectrum. Moreover, there is evidence suggesting that after a long transient regime, characterized by an excessive production of small loops at the scale of initial conditions, the characteristic size of large loops starts to grow linearly with time \cite{NewLoops}. (See also \cite{NewOthers} for the numerical results and interpretations reported by other groups.) Unfortunately, the dynamical range of these simulations is not sufficiently large to confirm the numerical results with a high confidence level. 

In addition to numerical studies, a number of analytical models were proposed \cite{AnalyticKibble, NewAnalytic}, but the downside of most models is that they contain phenomenological parameters, such as the mean velocity of strings.  To overcome these problems and to better understand various nonlinear stringy phenomena, such as cross correlations \cite{Crosscorrelations} and semi-scaling \cite{Semiscaling}, we developed a kinetic theory of strings \cite{KineticTheory}. The stochastic dynamics of strings was  modeled with a transport equation which is capable of describing long strings with Nambu-Goto evolution as well as interactions taken into account; and the corresponding system of coupled transport equations which can simultaneously describe long strings as well as closed loops. In this paper we extend the kinetic theory approach further and derive a fully hydrodynamic description of strings.

The paper is organized as follows. In the next section we derive a simplified version of the transport equation for strings. In Sec. \ref{Equilibrium} we prove an $H$-theorem and derive an equilibrium condition for strings. In Sec. \ref{Conservation} we prove a conservation theorem and obtain a general set of conservation equations. The hydrodynamic equations for von Mises-Fisher distributions are derived in Sec. \ref{Hydrodynamics} under assumption of a local thermodynamic equilibrium. In Sec. \ref{Nonequilibrium} we relax the equilibrium assumption and discuss the consequences of nonequilibrium steady states. The main results of the paper are summarized in Sec.  \ref{summary}. 

\section{Transport Equation}\label{TransportEquation}

The dynamics of cosmic strings without interactions, such as reconnections of nearby strings or production of string loops, is described fairly well by Nambu-Goto action,
\be
S_{NG} = - \mu \int \sqrt{-{\det}(h_{ab})} d\sigma dt  
\ee
where $\mu$ is the string tension and
\be
h_{ab} = g_{\mu\nu} x^\mu_{,a}x^\nu_{,b}.
\ee
 The corresponding equation of motion is
\be
x_{,a}^{\mu\, ;a} + \Gamma^{\mu}_{\nu \tau} h^{ab} x^\nu_{,a}x^\tau_{,b} =0.
\label{eq:EOM}
\ee
After gauge fixing ($h_{01} =0$, $h_{00}+h_{11}=0$, $t=x^0$) on a flat background we get a system of equations for a three-vector, $\bf x$,
\bea
\frac{d^2 {\bf x}}{dt^2} - \frac{d^2 {\bf x}}{d\sigma^2} = 0, \label{eq:wave_equation}\\
\left(\frac{d{\bf x}}{dt}\right)^2 + \left(\frac{d {\bf x}}{d\sigma}\right)^2 = 1, \label{eq:constraint1}\\
\frac{d{\bf x}}{d\sigma} \cdot \frac{d{\bf x}}{dt} = 0, \label{eq:constraint2}
\eea
where the speed of light is set to one. The solution of  (\ref{eq:wave_equation}) can be described in terms of right-moving ${\bf a}(\sigma-t)$ and left-moving ${\bf b}(\sigma+t)$ waves
\be
\bold{x}(\sigma, t) = \frac{\bold{a}(\sigma- t) +\bold{b}(\sigma + t) }{2}
\label{eq:wave}
\ee
with additional constraints, 
\be
\left |\frac{d{\bf a}}{d\sigma} \right |=\left |\frac{d{\bf b}}{d\sigma} \right |=1.
\ee
due to (\ref{eq:constraint1}) and (\ref{eq:constraint2}).

In general the exact evolution of a given infinitesimal segment of string is known only if all of the functions, $\bold{a}(\sigma)$ and $\bold{b}(\sigma)$, for all strings are known. In the kinetic theory of strings such evolution is described by a system of equations analogous to the Bogoliubov-Born-Green-Kirkwood-Yvon hierarchy for particles  \cite{KineticTheory}. Truncating the hierarchy at a given order is equivalent to neglecting higher-order effects and is often a starting point in a derivation of a transport equation. The first attempt to drive a transport equation for strings was given in Ref.  \cite{KineticTheory} where the state of the system was modeled with a $9+1$ dimensional distribution function $f({\bf A}, {\bf B}, {\bf x}, t)$ of the correlation vectors ${\bf A}$ and ${\bf B}$, comoving position ${\bf x}$, and conformal time $t$. In this paper we make a crucial simplification by concentrating instead on a $7+1$ dimensional distribution $f(\hat{\bf A}, \hat{\bf B}, {\bf x}, t)$ where
\be
\hat{\bf A}\equiv - \frac{d{\bf a}}{d\sigma} \;\;\;\;\;\text{and}\;\;\;\;\; \hat{\bf B}\equiv \frac{d{\bf b}}{d\sigma} 
\ee
are the unit three-vectors corresponding to tangent vectors of right- and left-moving waves.  

In a Friedmann universe described by a Friedmann-Robertson-Walker metric,
\be
ds^2 = a(t)^2 (dt^2 - d{\bf x}^2),
\ee
one can show that the unit vectors evolve according to 
\bea
\frac{d }{dt} \hat{\bf A} = -{\cal H} (\hat{\bf B} - (\hat{\bf A} \cdot \hat{\bf B}) \hat{\bf A}) \label{eq:dA}\\
\frac{d }{dt} \hat{\bf B} = -{\cal H} (\hat{\bf A} - (\hat{\bf A} \cdot \hat{\bf B}) \hat{\bf B}) \label{eq:dB}
\eea
and the energy density decays as,
\be
\frac{d }{dt} f =- {\cal H}  \frac{ 1}{2} \left (\hat{\bf A}  + \hat{\bf B} \right )^2  {f} \label{eq:df},
\ee
where $H\equiv \frac{\dot{a}}{a}$ is the Hubble rate in conformal coordinates (See, for example, Ref. \cite{BB}). Then one can derive a transport equation for  $f(\hat{\bf A}, \hat{\bf B}, {\bf x}, t)$ by following the evolution of $f(\hat{\bf A}, \hat{\bf B}, {\bf x}, t)$ from time $t$ to time $t+\delta t$. For an infinitesimally small $\delta t$, we can use (\ref{eq:dA}), (\ref{eq:dB}) and (\ref{eq:df}) to express the distribution at time $t+\delta t$ in terms of the distribution at time $t$,
\bea
f\left (\hat{\bf A}- {\cal H}  \left( \hat{\bf B} - \left (\hat{\bf A}\cdot \hat{\bf B} \right ) \hat{\bf A}\right ) \delta t, \hat{\bf B} - {\cal H}  \left( \hat{\bf A} - \left (\hat{\bf A}\cdot \hat{\bf B} \right ) \hat{\bf B}\right ) \delta t, {\bf x} + \frac{ 1}{2} \left (\hat{\bf A}  + \hat{\bf B} \right ) \delta t, t + \delta t \right )= \notag\\ 
\left (1-{\cal H}  \frac{1}{2} \left (\hat{\bf A}  + \hat{\bf B} \right )^2 \right) f(\hat{\bf A}, \hat{\bf B}, {\bf x}, t)  + \left ( \frac{\partial f}{\partial t} \right )_{coll} \delta t.  \;\;\;\;\;\;\;\;\;\;\; \;\;\;\;\;
\eea
The ``collision'' term on the right-hand side represents the Nambu-Goto evolution and interactions. By expanding to the linear order in $\delta t$, we obtain the following equation
\bea
\left ( \frac{\partial}{\partial t} + \frac{ 1}{2} \left (\hat{\bf A}  + \hat{\bf B} \right ) \cdot \frac{\partial}{\partial {\bf x}} + {\cal H} \left ( \frac{1}{2} \left (\hat{\bf A}  + \hat{\bf B} \right )^2 - \hat{\bf A} \cdot \frac{\partial}{\partial \hat{\bf B}} - \hat{\bf B} \cdot \frac{\partial}{\partial \hat{\bf A}} \right )  \right) f(\hat{\bf A}, \hat{\bf B})   = \left ( \frac{\partial f}{\partial t} \right )_{coll},\;\; \;\;\;\;\; \;\;\;\;\;
\label{eq:transport0}
\eea
where the dependence on the comoving position ${\bf x}$ and conformal time $t$ is suppressed for brevity of notations.  Note, that the use of the simplified distribution function $f(\hat{\bf A}, \hat{\bf B}, {\bf x}, t)$ does not necessarily imply that the correlations must be small. In fact, the long-distance correlations may still exist whenever the distributions of $\hat{\bf A}$ and $\hat{\bf B}$ are peaked sharply around a particular direction. 

The effects of ``collisions'' can be estimated as in Ref. \cite{KineticTheory}. The Nambu-Goto ``collisions'' proceed with the same rate, regardless of the orientations of string world sheets, as the right- and left-moving waves pass through each other.  The rate of such ``collisions'' is given by
\be
\Gamma_\text{evolution} \sim  \frac{1}{l_\text{min}},
 \label{eq:evolution_rate}
\ee
where $l_\text{min}$  describes  the resolution scale of the kinetic theory. Since the Nambu-Goto action is only accurate on sufficiently large scales there is a lower bond $l_{\min} \lesssim \mu^{-1/2}$. However, on the cosmological settings the resolution scale might also be set by the scale of the initial conditions or by the scale of the gravitation backreaction. The exact value of $l_\text{min}$ is unimportant for the subsequent discussion. 

The rate of the second kind of ``collision'', due to interactions between a segment described by tangent vectors $\hat{\bf A}$ and $\hat{\bf B}$ with another segment described by tangent vectors $\hat{\bf A}'$ and $\hat{\bf B}'$, is given by a probability of intersecting in the target space. In a universe with three large spatial dimensions the probability must be proportional to the local energy density, $\rho$,  the intercommutation probability, $p$,\footnote{The intercommutation probability is a free parameter of order one for the field theory strings but can also be much smaller than one of the string theory strings.} and also to what we call a cross-sectional volume. Intuitively, the cross-sectional volume describes a likelihood of interactions of string segments due to relative orientations of their world sheets $(1, \hat{\bf A}) \wedge (1, \hat{\bf B})$ and $(1, \hat{\bf A}') \wedge (1, \hat{\bf B}')$. Altogether we get the following expression for the rate of interactions,
\be
\Gamma_\text{interaction} \sim \gamma \rho \;p \; \left | (1, \hat{\bf A}) \wedge (1, \hat{\bf B}) \wedge (1, \hat{\bf A}') \wedge   (1, \hat{\bf B}')   \right |
 \label{eq:interaction_rate}
\ee 
where
\be
\rho  \equiv \int \text{d} \hat{\bf A} \text{d} \hat{\bf B}  f(\hat{\bf A}, \hat{\bf B})
\ee 
and $\gamma$ is some constant.

When a ``collision'' takes place the corresponding opposite-moving tangent vectors are interchanged, e.g.
\be
 (\hat{\bf A}', \hat{\bf B})\;\;\text{and}\;\;(\hat{\bf A}, \hat{\bf B}') \;\;\;\;\;\;\Rightarrow \;\;\;\;\;\;(\hat{\bf A}, \hat{\bf B})\;\;\text{and}\;\;(\hat{\bf A}', \hat{\bf B}'),
\ee
or
\be
 (\hat{\bf A}, \hat{\bf B})\;\;\text{and}\;\;(\hat{\bf A}', \hat{\bf B}') \;\;\;\;\;\;\Rightarrow \;\;\;\;\;\;(\hat{\bf A}', \hat{\bf B})\;\;\text{and}\;\;(\hat{\bf A}, \hat{\bf B}').
\ee
This gives rise to a couple of integral terms which can be combined together to yield the following expression,
\be
\left ( \frac{\partial f}{\partial t} \right )_{coll} =\int \text{d} \hat{\bf A}' \text{d} \hat{\bf B}' \left (\Gamma_\text{evolution} +\Gamma_\text{interaction} \right ) \frac{ f(\hat{\bf A}', \hat{\bf B}) f(\hat{\bf A}, \hat{\bf B}')- f(\hat{\bf A}, \hat{\bf B}) f(\hat{\bf A}', \hat{\bf B}')}{\rho}
\label{eq:collision}
\ee
By substituting (\ref{eq:collision}) into (\ref{eq:transport0}) we obtain the final form of the transport equation, 
\bea
\left ( \frac{\partial}{\partial t} + \frac{ 1}{2} \left (\hat{\bf A}  + \hat{\bf B} \right ) \cdot \frac{\partial}{\partial {\bf x}} + {\cal H} \left ( \frac{1}{2} \left (\hat{\bf A}  + \hat{\bf B} \right )^2 - \hat{\bf A} \cdot \frac{\partial}{\partial \hat{\bf B}} - \hat{\bf B} \cdot \frac{\partial}{\partial \hat{\bf A}} \right )  \right) f(\hat{\bf A}, \hat{\bf B})   = \notag\\ \int \text{d} \hat{\bf A}' \text{d} \hat{\bf B}' \;{\cal P}(\hat{\bf A},\hat{\bf B},\hat{\bf A}',\hat{\bf B}') \left (  f(\hat{\bf A}', \hat{\bf B}) f(\hat{\bf A}, \hat{\bf B}')- f(\hat{\bf A}, \hat{\bf B}) f(\hat{\bf A}', \hat{\bf B}')\right )
\label{eq:transport}
\eea
where
\bea
{\cal P}(\hat{\bf A},\hat{\bf B},\hat{\bf A}',\hat{\bf B}') \sim \frac{1}{l_\text{min}\,\rho} +\gamma\; p \left | (1, \hat{\bf A}) \wedge (1, \hat{\bf B}) \wedge (1, \hat{\bf A}') \wedge   (1, \hat{\bf B}')   \right |. 
\label{eq:scatter}
\eea
Note that the function ${\cal P}(\hat{\bf A},\hat{\bf B},\hat{\bf A}',\hat{\bf B}')$ is invariant under permutations of variables. This property will turn out to be crucial for proving an $H$-theorem and a conservation theorem in the following sections.

\section{Equilibrium Distributions}\label{Equilibrium}

Consider a localized version of the transport equation (\ref{eq:transport}) where the spatial inhomogeneities and gravitational effects are ignored, i.e.
\be
\frac{\partial}{\partial t} f({\bf A}, {\bf B})= \int \text{d} \hat{\bf A}' \text{d} \hat{\bf B}' \;{\cal P}(\hat{\bf A},\hat{\bf B},\hat{\bf A}',\hat{\bf B}') \left (  f(\hat{\bf A}', \hat{\bf B}) f(\hat{\bf A}, \hat{\bf B}')- f(\hat{\bf A}, \hat{\bf B}) f(\hat{\bf A}', \hat{\bf B}')\right ).
\label{eq:H0}
\ee
If we assume that a local equilibrium (defined by $\frac{\partial}{\partial t} f_{eq}({\bf A}, {\bf B})=0$) is established much faster than all other time scales, then the right-hand side of (\ref{eq:H0}) must also vanish,
\be
\int \text{d} \hat{\bf A}' \text{d} \hat{\bf B}' \;{\cal P}(\hat{\bf A},\hat{\bf B},\hat{\bf A}',\hat{\bf B}') \left (  f(\hat{\bf A}', \hat{\bf B}) f(\hat{\bf A}, \hat{\bf B}')- f(\hat{\bf A}, \hat{\bf B}) f(\hat{\bf A}', \hat{\bf B}')\right )=0.
\ee 
We are now ready to prove an $H$-theorem for strings which is analogous to the famous Boltzmann's $H$-theorem for particles.  In particular, we will show that
\be
H({\bf x},t) \equiv \int \text{d} \hat{\bf A} \text{d} \hat{\bf B}  f(\hat{\bf A}, \hat{\bf B}) \log( f(\hat{\bf A}, \hat{\bf B})),
\label{eq:Hfunction}
\ee
can never increase with time
\be
\frac{d H}{dt} \le 0,
\label{eq:Htheorem}
\ee
whenever $f(\hat{\bf A}, \hat{\bf B})$ satisfies the local transport equation (\ref{eq:H0}). 

By differentiating (\ref{eq:Hfunction}) with respect to time, we obtain
\be
\frac{d H}{dt} =    \int \text{d} \hat{\bf A} \text{d} \hat{\bf B}  \frac{\partial  f(\hat{\bf A}, \hat{\bf B})}{\partial t} \left (1+\log( f(\hat{\bf A}, \hat{\bf B})) \right) \label{eq:dHdt}
\ee
and after substituting (\ref{eq:H0}) into (\ref{eq:dHdt}) we arrive at the following equation
\bea
\frac{d H}{dt} =    \int \text{d} \hat{\bf A} \text{d} \hat{\bf B}  \text{d} \hat{\bf A}' \text{d} \hat{\bf B}' \;{\cal P}(\hat{\bf A},\hat{\bf B},\hat{\bf A}',\hat{\bf B}') \left (  f(\hat{\bf A}', \hat{\bf B}) f(\hat{\bf A}, \hat{\bf B}')- f(\hat{\bf A}, \hat{\bf B}) f(\hat{\bf A}', \hat{\bf B}')\right ) \notag \times \\  \left (1+\log( f(\hat{\bf A}, \hat{\bf B})) \right).
\label{eq:H1}
\eea
Since ${\cal P}(\hat{\bf A},\hat{\bf B},\hat{\bf A}',\hat{\bf B}')$ is invariant under permutations of variables, the right-hand side of (\ref{eq:H1}) can be rewritten by interchanging the dummy variables of integration $\hat{\bf A}, \hat{\bf B}$ and $\hat{\bf A}', \hat{\bf B}'$,
\bea
\frac{d H}{dt} =    \int \text{d} \hat{\bf A}' \text{d} \hat{\bf B}'  \text{d} \hat{\bf A} \text{d} \hat{\bf B} \;{\cal P}(\hat{\bf A}',\hat{\bf B}',\hat{\bf A},\hat{\bf B}) \left (  f(\hat{\bf A}, \hat{\bf B}') f(\hat{\bf A}', \hat{\bf B})- f(\hat{\bf A}', \hat{\bf B}') f(\hat{\bf A}, \hat{\bf B})\right ) \notag \times \\  \left (1+\log( f(\hat{\bf A}', \hat{\bf B}')) \right).
\label{eq:H2}
\eea
Note that (\ref{eq:H1}) and (\ref{eq:H2}) only differ in an argument of the logarithms. Thus, we can add (\ref{eq:H1}) and (\ref{eq:H2}) to get
\bea
2\frac{d H}{dt} =    \int \text{d} \hat{\bf A}' \text{d} \hat{\bf B}'  \text{d} \hat{\bf A} \text{d} \hat{\bf B} \;{\cal P}(\hat{\bf A}',\hat{\bf B}',\hat{\bf A},\hat{\bf B}) \left (  f(\hat{\bf A}, \hat{\bf B}') f(\hat{\bf A}', \hat{\bf B})- f(\hat{\bf A}', \hat{\bf B}') f(\hat{\bf A}, \hat{\bf B})\right ) \notag \times \\  \left (2+\log( f(\hat{\bf A}, \hat{\bf B}) f(\hat{\bf A}', \hat{\bf B}')) \right)
\label{eq:H3}
\eea
and by interchanging $\hat{\bf A}$ and $ \hat{\bf A}'$ we obtain another equivalent equation,
\bea
2\frac{d H}{dt} =    \int \text{d} \hat{\bf A}' \text{d} \hat{\bf B}'  \text{d} \hat{\bf A} \text{d} \hat{\bf B} \;{\cal P}(\hat{\bf A}',\hat{\bf B}',\hat{\bf A},\hat{\bf B}) \left (  f(\hat{\bf A}', \hat{\bf B}') f(\hat{\bf A}, \hat{\bf B})- f(\hat{\bf A}, \hat{\bf B}') f(\hat{\bf A}', \hat{\bf B})\right ) \notag \times \\  \left (2+\log( f(\hat{\bf A}', \hat{\bf B}) f(\hat{\bf A}, \hat{\bf B}')) \right).
\label{eq:H4}
\eea
To arrive at the final expression we add (\ref{eq:H3}) and (\ref{eq:H4}),
\bea
\frac{dH}{d t}= - \frac{1}{4} \int \text{d} \hat{\bf A}' \text{d} \hat{\bf B}'  \text{d} \hat{\bf A} \text{d} \hat{\bf B} \;{\cal P}(\hat{\bf A}',\hat{\bf B}',\hat{\bf A},\hat{\bf B}) \left (  f(\hat{\bf A}, \hat{\bf B}') f(\hat{\bf A}', \hat{\bf B})- f(\hat{\bf A}', \hat{\bf B}') f(\hat{\bf A}, \hat{\bf B})\right ) \notag \times \\  \left (\log( f(\hat{\bf A}, \hat{\bf B}') f(\hat{\bf A}', \hat{\bf B})) -  \log( f(\hat{\bf A}, \hat{\bf B}) f(\hat{\bf A}', \hat{\bf B}')) \right).\;\;\;\;\;
\label{eq:H5}
\eea
But since $\log$ is a monotonically increasing function we conclude that 
\be
\left (  f(\hat{\bf A}, \hat{\bf B}') f(\hat{\bf A}', \hat{\bf B})- f(\hat{\bf A}', \hat{\bf B}') f(\hat{\bf A}, \hat{\bf B})\right ) \left (\log( f(\hat{\bf A}, \hat{\bf B}') f(\hat{\bf A}', \hat{\bf B})) -  \log( f(\hat{\bf A}, \hat{\bf B}) f(\hat{\bf A}', \hat{\bf B}')) \right) \ge 0 \notag
\label{eq:H6}
\ee
as well as $\frac{dH}{dt}\le0$, which is exactly the statement of the $H$-theorem for strings (\ref{eq:Htheorem}). Note that the equality $\frac{dH}{dt}=0$ is realized if and only if 
\be
f_{eq}(\hat{\bf A}, \hat{\bf B}') f_{eq}(\hat{\bf A}', \hat{\bf B}) = f_{eq}(\hat{\bf A}', \hat{\bf B}') f_{eq}(\hat{\bf A}, \hat{\bf B})
 \label{eq:H8}
\ee
 corresponding to an equilibrium distribution $f_{eq}(\hat{\bf A}, \hat{\bf B})$.
 
In addition, we can show that for an arbitrary equilibrium distribution, there exists a factorization
\be
 f_{eq}(\hat{\bf A}, \hat{\bf B}) = \rho\; p_a(\hat{\bf A})\; p_b(\hat{\bf B}) 
\label{eq:decomposition}
\ee
where  $p_a(\hat{\bf A})$ and $p_b(\hat{\bf B})$ are the normalized probability distributions, i.e.
\be
\int \text{d} \hat{\bf A} \; p_a(\hat{\bf A}) = \int \text{d} \hat{\bf B} \; p_b(\hat{\bf B}) =1.
\ee
Indeed, for an arbitrary pair of unit vectors $\hat{\bf A}''$ and $\hat{\bf B}''$ such that $ f(\hat{\bf A}'', \hat{\bf B}'') \neq 0$, we can define
\be
p_a(\hat{\bf A})  \equiv  \frac{f_{eq}(\hat{\bf A}, \hat{\bf B}'')}{\int \text{d} \hat{\bf A}'  f_{eq}(\hat{\bf A}', \hat{\bf B}'')  } \;\;\; \text{and}\;\;\; p_b(\hat{\bf B})  = \frac{f_{eq}(\hat{\bf A}'', \hat{\bf B})}{\int \text{d} \hat{\bf B}'  f_{eq}(\hat{\bf A}'', \hat{\bf B}')  }.
\ee
Then it is easy to check that the condition (\ref{eq:H8}) implies (\ref{eq:decomposition}),
\bea
f_{eq}(\hat{\bf A}, \hat{\bf B}') f_{eq}(\hat{\bf A}', \hat{\bf B}) = f_{eq}(\hat{\bf A}', \hat{\bf B}') f_{eq}(\hat{\bf A}, \hat{\bf B})\;\;\;\; \;\; \forall \;\;\;\hat{\bf A},\hat{\bf B}, \hat{\bf A}', \hat{\bf B}'  \Rightarrow \notag\\
\rho \; p_a(\hat{\bf A})  p_b(\hat{\bf B}) = \frac{ \int \text{d} \hat{\bf A}' \text{d} \hat{\bf B}'  f_{eq}(\hat{\bf A}', \hat{\bf B}')  f_{eq}(\hat{\bf A}, \hat{\bf B}'') f_{eq}(\hat{\bf A}'', \hat{\bf B})}{\int \text{d} \hat{\bf A}'  f_{eq}(\hat{\bf A}', \hat{\bf B}'')\int \text{d} \hat{\bf B}'  f_{eq}(\hat{\bf A}'', \hat{\bf B}')} =\;\;\;\;\;\;\;\;\;\;\;\;\;\;\;\;\;\\ \frac{\int \text{d} \hat{\bf A}' \text{d} \hat{\bf B}'  f_{eq}(\hat{\bf A}', \hat{\bf B}')   f_{eq}(\hat{\bf A}'', \hat{\bf B}'')  f_{eq}(\hat{\bf A}, \hat{\bf B})}{\int \text{d} \hat{\bf A}' \text{d} \hat{\bf B}'  f_{eq}(\hat{\bf A}', \hat{\bf B}')  f_{eq}(\hat{\bf A}'', \hat{\bf B}'')} =  f_{eq}(\hat{\bf A}, \hat{\bf B})  \notag
\eea
Of course, the opposite is also true, and (\ref{eq:decomposition}) implies (\ref{eq:H8}). Thus, we have four equivalent definitions of a local equilibrium of strings
\bea
\frac{\partial}{\partial t} f_{eq}(\hat{\bf A}, \hat{\bf B})  = 0  \Leftrightarrow  \frac{d}{d t} H[f_{eq}(\hat{\bf A}, \hat{\bf B})]  = 0   \Leftrightarrow  f_{eq}(\hat{\bf A}, \hat{\bf B}) = \rho\; p_a(\hat{\bf A})\; p_b(\hat{\bf B}) \Leftrightarrow \notag \\
f_{eq}(\hat{\bf A}, \hat{\bf B}') f_{eq}(\hat{\bf A}', \hat{\bf B}) = f_{eq}(\hat{\bf A}', \hat{\bf B}') f_{eq}(\hat{\bf A}, \hat{\bf B})\;\;\;\; \;\; \forall \;\;\;\hat{\bf A},\hat{\bf B}, \hat{\bf A}', \hat{\bf B}'. \label{eq:equilibrium}
\eea

\section{Conservation Theorem}\label{Conservation}

Consider an arbitrary quantity $Q(\hat{\bf A}, \hat{\bf B})$ which is conserved during ``collisions'' (Nambu-Goto or interactions), i.e. 
\be
Q(\hat{\bf A}, \hat{\bf B})  +Q(\hat{\bf A}', \hat{\bf B}') =Q(\hat{\bf A}', \hat{\bf B}) + Q(\hat{\bf A}, \hat{\bf B}') .
\label{eq:conservation1}
\ee 
Then we can integrate both sides of the transport equation (\ref{eq:transport}) to get 
\bea
 \int \text{d} \hat{\bf A} \text{d} \hat{\bf B} \;Q(\hat{\bf A}, \hat{\bf B}) \left ( \frac{\partial}{\partial t} + \frac{ 1}{2} \left (\hat{\bf A}  + \hat{\bf B} \right ) \cdot \frac{\partial}{\partial {\bf x}} + {\cal H} \left ( \frac{1}{2} \left (\hat{\bf A}  + \hat{\bf B} \right )^2 - \hat{\bf A} \cdot \frac{\partial}{\partial \hat{\bf B}} - \hat{\bf B} \cdot \frac{\partial}{\partial \hat{\bf A}} \right )  \right) f(\hat{\bf A}, \hat{\bf B})    = \notag\\
 \int \text{d} \hat{\bf A} \text{d} \hat{\bf B} \text{d} \hat{\bf A}' \text{d} \hat{\bf B}' \;Q(\hat{\bf A}, \hat{\bf B})   \;{\cal P}(\hat{\bf A},\hat{\bf B},\hat{\bf A}',\hat{\bf B}') \left (  f(\hat{\bf A}', \hat{\bf B}) f(\hat{\bf A}, \hat{\bf B}')- f(\hat{\bf A}, \hat{\bf B}) f(\hat{\bf A}', \hat{\bf B}')\right ), \;\;\;\;\;\;\;
 \label{eq:conservation2}
\eea
where the right-hand side can be rewritten as
\bea
 \int \text{d} \hat{\bf A} \text{d} \hat{\bf B}  \text{d} \hat{\bf A}' \text{d} \hat{\bf B}'  \;Q(\hat{\bf A}, \hat{\bf B}) \;{\cal P}(\hat{\bf A},\hat{\bf B},\hat{\bf A}',\hat{\bf B}') \left (  f(\hat{\bf A}', \hat{\bf B}) f(\hat{\bf A}, \hat{\bf B}')- f(\hat{\bf A}, \hat{\bf B}) f(\hat{\bf A}', \hat{\bf B}')\right ) & = \notag \\
 \int \text{d} \hat{\bf A} \text{d} \hat{\bf B}   \text{d} \hat{\bf A} '\text{d} \hat{\bf B}' \;Q(\hat{\bf A}', \hat{\bf B}) \;{\cal P}(\hat{\bf A},\hat{\bf B},\hat{\bf A}',\hat{\bf B}') \left (  f(\hat{\bf A}, \hat{\bf B}) f(\hat{\bf A}', \hat{\bf B}')- f(\hat{\bf A}', \hat{\bf B}) f(\hat{\bf A}, \hat{\bf B}')\right ) & = \notag \\
 \int \text{d} \hat{\bf A} \text{d} \hat{\bf B}  \text{d} \hat{\bf A}' \text{d} \hat{\bf B}' \;Q(\hat{\bf A}, \hat{\bf B}')  \;{\cal P}(\hat{\bf A},\hat{\bf B},\hat{\bf A}',\hat{\bf B}') \left (  f(\hat{\bf A}', \hat{\bf B}') f(\hat{\bf A}, \hat{\bf B})- f(\hat{\bf A}, \hat{\bf B}') f(\hat{\bf A}', \hat{\bf B})\right ) & = \notag \\
 \int \text{d} \hat{\bf A} \text{d} \hat{\bf B}   \text{d} \hat{\bf A}' \text{d} \hat{\bf B}' \;Q(\hat{\bf A}', \hat{\bf B}') \;{\cal P}(\hat{\bf A},\hat{\bf B},\hat{\bf A}',\hat{\bf B}') \left (  f(\hat{\bf A}, \hat{\bf B}') f(\hat{\bf A}', \hat{\bf B})- f(\hat{\bf A}', \hat{\bf B}') f(\hat{\bf A}, \hat{\bf B})\right ) & \notag
\eea
due to invariance of ${\cal P}(\hat{\bf A},\hat{\bf B},\hat{\bf A}',\hat{\bf B}')$ under permutations of variables.  By adding these four equivalent expressions together  and dividing by four we obtain yet another equivalent expression
\bea
\frac{1}{4} \int \text{d} \hat{\bf A} \text{d} \hat{\bf B} \; \text{d} \hat{\bf A}' \text{d} \hat{\bf B}' \;{\cal P}(\hat{\bf A},\hat{\bf B},\hat{\bf A}',\hat{\bf B}') \left (  f(\hat{\bf A}', \hat{\bf B}) f(\hat{\bf A}, \hat{\bf B}')- f(\hat{\bf A}, \hat{\bf B}) f(\hat{\bf A}', \hat{\bf B}')\right )  \times \notag \\ \left ( Q(\hat{\bf A}, \hat{\bf B})  + Q(\hat{\bf A}', \hat{\bf B}')  - Q(\hat{\bf A}', \hat{\bf B})  -Q(\hat{\bf A}, \hat{\bf B}') \right) \;\;\;\;\;\;\;\;\;\;\;
\label{eq:conservation3}
\eea
which must vanish for any conserved quantity  $Q(\hat{\bf A}, \hat{\bf B})$ due to (\ref{eq:conservation1}). Thus, the right-hand side of (\ref{eq:conservation2}) is identically zero and we arrive at the following equation,
\bea
 \int \text{d} \hat{\bf A} \text{d} \hat{\bf B} \;Q(\hat{\bf A}, \hat{\bf B}) \left ( \frac{\partial}{\partial t} + \frac{ 1}{2} \left (\hat{\bf A}  + \hat{\bf B} \right ) \cdot \frac{\partial}{\partial {\bf x}} + {\cal H} \left ( \frac{1}{2} \left (\hat{\bf A}  + \hat{\bf B} \right )^2 - \hat{\bf A} \cdot \frac{\partial}{\partial \hat{\bf B}} - \hat{\bf B} \cdot \frac{\partial}{\partial \hat{\bf A}} \right )
 \right) f(\hat{\bf A}, \hat{\bf B})    =  0.\;\;\;\;\;\;
 \label{eq:conservation}
\eea

It is now convenient to define an expectation value
\be
\langle O \rangle \equiv \frac{1}{\rho} \int \text{d}  \hat{\bf A} \text{d} \hat{\bf B} \; O(\hat{\bf A}, \hat{\bf B})  f(\hat{\bf A}, \hat{\bf B}). 
\ee
and to rewrite (\ref{eq:conservation}), after integrating by parts, as
\bea
\frac{\partial }{\partial t}  \left  \langle  \rho Q \right  \rangle + \frac{1}{2} \frac{\partial}{\partial x_j}  \left \langle  \rho \left ( A_j + B_j \right ) Q \right \rangle +  {\cal H} \left \langle \rho \left ( \frac{1}{2} \left (A_j + B_j \right )^2 + A_j \frac{\partial}{\partial B_j}+ B_j \frac{\partial}{\partial A_j}\right) Q \right \rangle = 0,\;\;
\label{eq:conservation_theorem}
\eea
where the summation over repeated indices is implied.  In the literature on the kinetic theory of particles a similar expression is known as a conservation theorem. By applying the conservation theorem to different (locally) conserved quantities we can derive different conservation equations.

In particular, an equation for a local conservation of energy is obtained by setting $Q=1$ in (\ref{eq:conservation_theorem}), 
\bea
\frac{\partial }{\partial t} \rho + \frac{1}{2} \frac{\partial}{\partial x_j} \rho  \left \langle A_j + B_j  \right \rangle+ {\cal H}  \rho  \left \langle 1+ A_j B_j \right \rangle=0.
\label{eq:one}
\eea
Another important example is a local conservation of right-moving and left-moving tangent vectors. This gives us six more equations, one for each of the components of each of the tangent vector. By setting  $Q$ to $A_1, A_2, A_3, B_1, B_2$ or $B_3$ in  (\ref{eq:conservation_theorem}), we obtain the following equations
\bea
\frac{\partial }{\partial t}  \rho \left  \langle A_i \right  \rangle + \frac{1}{2} \frac{\partial}{\partial x_j}  \rho \left \langle A_j A_i + B_j A_i \right \rangle + {\cal H}  \rho  \left \langle A_j B_j A_i + A_i+B_i \right \rangle =0,
\label{eq:right}\\
\frac{\partial }{\partial t}  \rho  \left  \langle B_i \right  \rangle + \frac{1}{2} \frac{\partial}{\partial x_j}  \rho \left \langle A_j B_i + B_j B_i \right \rangle + {\cal H}  \rho  \left \langle A_j B_j B_i + A_i+B_i \right \rangle =0.
\label{eq:left}
\eea

Altogether, there are seven conservation equations [i.e. (\ref{eq:one}), (\ref{eq:right}) and (\ref{eq:right})],  but this is certainly not all there is [e.g. $Q(\hat{\bf A})=\delta(\hat{\bf A} - \hat{\bf A}')$ is conserved for an arbitrary $\hat{\bf A}'$]. In fact, there is an infinite number of conservation conditions that one could derive, and it may seem as if the system is overconstraint. (Note that there are only five conserved quantities in a kinetic theory of particles: mass, energy, and momentum, which give rise to five hydrodynamic equations: the continuity equation, heat equation and Navier-Stokes equation.) However, as we will see in the following section, seven is the exactly the number of equations needed to describe the hydrodynamics of strings to the leading order, and all other conservation conditions become relevant only at higher orders in perturbation theory.

\section{Hydrodynamic Equations}\label{Hydrodynamics}

In Sec. \ref{Equilibrium} we have shown that a local equilibrium is established if and only if $f(\hat{\bf A}, \hat{\bf B})$ factors into a product of the energy density, $\rho$, and two normalized probability distribution functions on a unit sphere, $p_a(\hat{\bf A})$ and $p_b(\hat{\bf B})$. Clearly, an arbitrary choice of the normalized distributions $p_a(\hat{\bf A})$ and $p_b(\hat{\bf B})$ would satisfy the equilibrium conditions (\ref{eq:equilibrium}) and additional considerations are needed to determine the most physical distributions. 

From the conservation of energy and of tangent vectors (of right- and left-moving waves) we derived equations (\ref{eq:right}), (\ref{eq:left}), and (\ref{eq:one}), which can be rewritten under assumption of a local equilibrium as
\bea
\frac{\partial }{\partial t} \rho  \left  \langle A_i \right  \rangle + \frac{1}{2} \frac{\partial}{\partial x_j} \rho \left ( \left \langle A_i A_j \right \rangle + \left \langle  A_i   \right \rangle \left \langle B_j \right \rangle \right)+ {\cal H}  \rho \left ( \left \langle A_j A_i \right \rangle \left \langle B_j \right \rangle  + \left \langle A_i\right \rangle+\left \langle B_i \right \rangle \right ) =0, \label{eq:hydrodynamic1}
\\
\frac{\partial }{\partial t} \rho  \left  \langle B_i \right  \rangle + \frac{1}{2} \frac{\partial}{\partial x_j} \rho \left ( \left \langle  B_i B_j \right \rangle + \left \langle B_i \right \rangle \left \langle  A_j \right \rangle \right )+ {\cal H}  \rho \left ( \left \langle B_j B_i \right \rangle \left \langle A_j \right \rangle  + \left \langle A_i\right \rangle+\left \langle B_i \right \rangle \right ) =0, \label{eq:hydrodynamic2}
\\
\frac{\partial }{\partial t} \rho + \frac{1}{2} \frac{\partial}{\partial x_j} \rho \left( \left \langle  A_j \right \rangle + \left \langle  B_j \right \rangle \right)+ {\cal H}  \rho  \left ( 1+ \left \langle A_j \right \rangle \left \langle B_j \right \rangle \right )=0.
\label{eq:hydrodynamic3}
\eea
This is a system of seven equations at the lowest order in perturbation theory, but for a self-consistent description, the energy distribution function $f(\hat{\bf A}, \hat{\bf B})$ must also be estimated with seven parameters. One parameter is the normalization constant, $\rho$, and the simplest three parameter family of normalized distributions is given by the von Mises-Fisher distributions on a unit sphere \cite{vMF},
\be
p_a(\hat{\bf A}) = \frac{1}{4 \pi \alpha \sinh(1/\alpha)} \exp\left( \frac{\hat{\bf u} \cdot \hat{\bf A}}{\alpha}\right ) \label{eq:pa}
\ee
and
\be
p_b(\hat{\bf B}) = \frac{1}{4 \pi \beta \sinh(1/\beta)} \exp\left( \frac{\hat{\bf v} \cdot \hat{\bf B}}{\beta}\right ) \label{eq:pb}
\ee
where $\hat{\bf u}$, $\hat{\bf v}$ are the mean directions and $\alpha, \beta$ are the dispersion parameters of right- and left-moving tangent vectors. The main motivation to use the von Mises-Fisher distributions is that they are mathematically tractable and at the same time give a good approximation of the Gaussian distributions wrapped on a sphere \cite{Breitenberger}. The corresponding lowest statistical momenta,
\bea
\langle A_i \rangle =   u_i F(\alpha) \label{eq:A},\\
\langle B_i \rangle =   v_i F(\beta) \label{eq:B},\\
\langle A_i A_j  \rangle =  \delta_{ij} \alpha F(\alpha)  + u_i u_j \left (1 - 3 \alpha F(\alpha) \right ) \label{eq:AA},
\eea
and
\bea
\langle B_i B_j  \rangle =  \delta_{ij} \beta F(\beta)   + u_i u_j \left (1 - 3 \beta F(\beta) \right ) \label{eq:BB} 
\eea
where, for convenience, we defined
\be
F(x) \equiv \coth\left(\frac{1}{x}\right ) - x
\ee
which can be approximated in the limit of small dispersions as
\be
F(x) \approx 1-x.
\ee

To obtain a complete system of hydrodynamic equations, we can plug the expressions (\ref{eq:A}),  (\ref{eq:B}),  (\ref{eq:AA}) and  (\ref{eq:BB}) into the equations (\ref{eq:hydrodynamic1}), (\ref{eq:hydrodynamic2}), (\ref{eq:hydrodynamic3}). The resulting equations are
\bea
\frac{\partial }{\partial t} \rho u_i F (\alpha) + \frac{1}{2} \frac{\partial}{\partial x_j} \rho\left (\delta_{ij} \alpha F (\alpha)   + u_i u_j \left (1 - 3 \alpha F (\alpha) \right ) + u_i v_j  F (\alpha)F (\beta) \right )+ \label{eq:hydrodynamicI}\\
{\cal H}  \rho \left (u_i F(\alpha) + v_i F(\beta) \left ( 1+ \alpha F(\alpha) \right ) + u_i u_j v_j  F(\beta) \left (1 - 3 \alpha F(\alpha) \right )   \right )=0, \notag \\
\frac{\partial }{\partial t} \rho v_i F(\beta) + \frac{1}{2} \frac{\partial}{\partial x_j} \rho\left (\delta_{ij} \beta F (\beta)   + v_i v_j \left (1 - 3 \beta F (\beta) \right ) ) +   v_i u_j F (\alpha)F (\beta)\right)+ \label{eq:hydrodynamicII} \\
{\cal H}  \rho \left (v_i F(\beta) + u_i F(\alpha) \left ( 1+ \beta F(\beta) \right ) + v_i  u_j v_j  F(\alpha) \left (1 - 3 \beta F(\beta) \right )   \right )=0, \notag
\eea
and
\bea
\frac{\partial }{\partial t} \rho + \frac{1}{2} \frac{\partial}{\partial x_j} \rho \left (  u_j F(\alpha) + v_j F(\beta) \right)+ {\cal H}  \rho  \left ( 1+u_i v_j  F (\alpha)F (\beta) \right )=0. \label{eq:hydrodynamicIII}
\eea
There are seven equations with seven unknowns: three scalar quantities, $\rho$, $\alpha$, $\beta$, and two vector quantities, $\hat{\bf u}$, $\hat{\bf v}$, constrained to unity, i.e., $|\hat{\bf u}|=|\hat{\bf v}|=1$. This system of partial differential equations is a huge simplification in comparison to the integro-differential transport equations of the kinetic theory (\ref{eq:transport}). 

Note that the hydrodynamic equations can be simplified further on a Minkowski background (i.e. ${\cal H}=0$), 
\bea
\frac{\partial }{\partial t} \rho u_i F (\alpha) + \frac{1}{2} \frac{\partial}{\partial x_j} \rho\left (\delta_{ij} \alpha F (\alpha)   + u_i u_j \left (1 - 3 \alpha F (\alpha) \right ) + u_i v_j  F (\alpha)F (\beta) \right )=0, \\
\frac{\partial }{\partial t} \rho v_i F(\beta) + \frac{1}{2} \frac{\partial}{\partial x_j} \rho\left (\delta_{ij} \beta F (\beta)   + v_i v_j \left (1 - 3 \beta F (\beta) \right ) ) +   v_i u_j F (\alpha)F (\beta)\right)=0, \\
\frac{\partial }{\partial t} \rho + \frac{1}{2} \frac{\partial}{\partial x_j} \rho \left (  u_j F(\alpha) + v_j F(\beta) \right)=0,
\eea
and in the limit of small dispersions $\alpha$ and $\beta$,
\bea
\frac{\partial }{\partial t} \rho u_i (1- \alpha) + \frac{1}{2} \frac{\partial}{\partial x_j} \rho\left (\delta_{ij} \alpha   + u_i u_j \left (1 - 3 \alpha \right ) + u_i v_j  (1 - \alpha - \beta) \right )=0, \label{eq:hydrodynamic_small1}\\
\frac{\partial }{\partial t} \rho v_i (1- \beta) + \frac{1}{2} \frac{\partial}{\partial x_j} \rho\left (\delta_{ij} \beta    + v_i v_j \left (1 - 3 \beta \right ) ) +   v_i u_j (1-\alpha-\beta)\right)=0, \label{eq:hydrodynamic_small2}\\
\frac{\partial }{\partial t} \rho + \frac{1}{2} \frac{\partial}{\partial x_j} \rho \left (  u_j (1-\alpha) + v_j (1-\beta) \right)=0.\label{eq:hydrodynamic_small3}
\eea
Evidently, even without gravitational effects, the evolution of right-moving waves, described by $u_i$ and $\alpha$, and left-moving waves, described by $v_i$ and $\beta$, is coupled not only through energy density $\rho$, but also directly though the cross terms 
\be
\frac{1}{2}  \frac{\partial}{\partial x_j} \rho v_i u_j F (\alpha)F (\beta)\;\;\;\text{and}\;\;\;\frac{1}{2} \frac{\partial}{\partial x_j} \rho u_i v_j F (\alpha)F (\beta).
\ee
These terms give rise to the so-called cross correlations between opposite moving waves  \cite{Crosscorrelations}, which are often overlooked in the literature on cosmic strings. Our analysis shows that the cross correlations appear at the leading order and should not be ignored. 

\section{Non-Equilibrium Steady States}\label{Nonequilibrium}

The reader might be puzzled by the fact that the hydrodynamic equations derived in the previous section do not depend on the intercommutation probability $p$. At first glance this seems very counterintuitive, given that there are numerical simulations that have demonstrated the dependence of string dynamics  on $p$, at least on small scales. Since the only assumption that we have made so far is the assumption of a local thermodynamic equilibrium, there must be a critical scale below which the assumption is no longer valid. The scale can be quite small for the field theory strings with $p\sim1$ but can also be rather large for the string theory strings with $p\ll 1$. 

Evidently, on small scales the network of strings must not behave as a closed Hamiltonian system whose equilibrium distribution would be given by the Liouville measure. For more general hyperbolic dynamical systems the nonequilibrium steady states provide a more accurate description of the dynamics.  These states were originally proposed by Sinai \cite{Sinai}, Ruelle \cite{Ruelle} and Bowen \cite{Bowen} and go by the name of the Sinai-Ruelle-Bowen (SRB) measures  \cite{Reviews}. In this paper we will refer to them as the SRB distributions.  Of course, an arbitrary local distribution can be substituted into (\ref{eq:one}), (\ref{eq:right}), and (\ref{eq:left}) to yield a new set of hydrodynamic equations for strings, and thus the main problem is to estimate the SRB distribution for strings. 

To derive a local distribution of strings we will apply the so-called thermodynamic formalism \cite{Reviews} with many ideas borrowed from the conventional statistical mechanics, but one important difference. In the statistical mechanics we are usually interested in the phase space states, $X$, when in the dynamical system of strings the key role is played by the phase space trajectories,  $X(t)$. Thus, it is convenient to think of time, $t$,  as a thermodynamical volume which is a conjugate variable to the topological pressure defined as
\be
p = \lim_{T\rightarrow\infty} \frac{1}{T}  \log {\cal Z},
\ee
where 
\be
{\cal Z} =  \int \text{d} X(0) \exp\left({ - \int_0^T  \sum_{\lambda_i>0} \chi_i(X(t))dt}\right) \label{eq:partition}
\ee
is the dynamical partition function \cite{Reviews}. The sum of local Lyapunov exponents $\chi_i$ (defined as a local rate of separation of the nearby trajectories) is taken over directions corresponding to only positive global Lyapunov exponents $\lambda_i$ (defined as a rate of separation of nearby trajectories in the limit of infinite times).

According to the partition function (\ref{eq:partition}) the trajectories with large Lyapunov exponents are exponentially suppressed. This effect  can be easily estimated for strings from an expected collision rate of a given configuration. In the transport equation (\ref{eq:transport}) there are two types of ``collisions,'' one due to Nambu-Goto evolution and another due to interactions. As was already stressed, the Nambu-Goto ``collisions'' take place with the same rate regardless of the trajectory and can always be factored out. In contrast, the rate of interactions depends on the trajectory in question. Such collisions give rise to an exponential sensitivity to initial conditions and contribute to the overall sum of the positive local Lyapunov exponents in (\ref{eq:partition}). Then the effect of interactions can be estimated from (\ref{eq:interaction_rate}) as
\be
f_{SRB}(\hat{\bf A}, \hat{\bf B}) \propto p_a(\hat{\bf A}) p_b(\hat{\bf B}) \exp\left(- \gamma \rho\;p\; \left | (1, \hat{\bf A}) \wedge (1, \hat{\bf B}) \wedge (1, \hat{\bf u}) \wedge   (1, \hat{\bf v})   \right | \mu^{-1/2} \right ) \label{eq:SRB}.
\ee
where $\gamma$ is some constant and $\mu^{-1/2}$ is the time scale of collisions. Note that, in general, one must solve for the mean directions (i.e. $\hat{\bf u} = \langle \hat{\bf A} \rangle$ and $\hat{\bf v} = \langle \hat{\bf B} \rangle$) to determine all of the parameters of the distribution.  However, if we are only interested in a leading-order effect of the interactions, then we can assume that the true SRB distribution is not too far from an equilibrium distribution described by (\ref{eq:pa}) and (\ref{eq:pb}). By treating the nonequilibrium effects as a small perturbation, we obtain
\be
f_{SRB}(\hat{\bf A}, \hat{\bf B}) \propto \exp\left(\frac{\hat{\bf u} \cdot \hat{\bf A}}{\alpha}+\frac{\hat{\bf v} \cdot \hat{\bf B}}{\beta}-  \gamma \rho\;p \;\left | (1, \hat{\bf A}) \wedge (1, \hat{\bf B}) \wedge (1, \hat{\bf u}) \wedge   (1, \hat{\bf v})   \right | \mu^{-1/2} \right ) \label{eq:SRB_simple}.
\ee
This distribution together with equations (\ref{eq:one}), (\ref{eq:right}) and (\ref{eq:left}) can, in principle, be used to determine an improved set of hydrodynamic equations.

\section{Summary}\label{summary}

The main objective of the paper was to develop a hydrodynamic approach to cosmic strings based on the ideas of the kinetic theory originally introduced in Ref. \cite{KineticTheory}. In what follows we will summarize the key  results. 

First of all, we derived a simplified version of the transport equation (\ref{eq:transport}) with Nambu-Goto evolution, interactions (such as reconnections of nearby strings and production of string loops) and background gravitational effects (such as Friedmann expansion)  taken into account. Secondly, we proved an $H$-theorem (\ref{eq:Htheorem}) and derived the four equivalent equilibrium conditions for strings (\ref{eq:equilibrium}). We also proved a conservation theorem (\ref{eq:conservation_theorem}) and derived a system of conservation equations  (\ref{eq:one}), (\ref{eq:right}) and (\ref{eq:left}).

Finally, we derived a complete set of hydrodynamic equations (\ref{eq:hydrodynamicI}), (\ref{eq:hydrodynamicII}) and (\ref{eq:hydrodynamicIII}) and discussed their possible generalizations in the context of the nonequilibrium steady states described by a Sinai-Ruelle-Bowen distribution (\ref{eq:SRB}). These equations provide an enormous simplification to the problem of string dynamics, but we will not attempt to solve them here. Instead, we will finish by noting that on a Minkowski background and in the limit of small dispersions $\alpha$ and $\beta$, the linearized equations take a particularly simple form (\ref{eq:hydrodynamic_small1}), (\ref{eq:hydrodynamic_small2}) and (\ref{eq:hydrodynamic_small3}). In fact, they are not much more complicated than the standard hydrodynamic equations for particles (i.e., continuity equation, heat equation, and Navier-Stokes equation), and there should not be any obstacles to solving these equations numerically, if not analytically.

\section*{Acknowledgments}

The author is grateful to Daniel Schubring for very helpful discussions and comments on the manuscript.

\end{document}